\documentclass[useAMS,usenatbib]{mn2e}
\usepackage{graphicx}

\usepackage[update,prepend]{epstopdf}

\title[Ion-proton polar caps]{A model of an ion-proton radio pulsar polar cap}
\author[P. B. Jones]{P. B. Jones\thanks{E-mail:
peter.jones@physics.ox.ac.uk}  \\
University of Oxford, Department of Physics, Denys Wilkinson Building,\\
Keble Road, Oxford OX1 3RH, U.K.}

\begin{document}

\date{}

\pagerange{\pageref{firstpage}--\pageref{lastpage}} \pubyear{}

\maketitle

\label{firstpage}

\begin{abstract}
A number of previous papers have developed an ion-proton theory of the pulsar polar cap.  The basic equations summarizing this are given here with the results of sets of model step-to-step calculations of pulse-precursor profiles.  The nature of step-to-step profile variations is described by calculated phase-resolved modulation indices. The conditions under which nulls are present in step sequences are analysed.  The change of mean null length with neutron-star surface temperature shows a pathway ending in emission similar to the Rotating Radio Transients.  The model accommodates exceptional pulsars, the millisecond pulsars (in principle), and the $8.5$ s period PSR J2144-3933.  These are considered separately and their emission mechanism discussed in some detail. 
\end{abstract}

\begin{keywords}
instabilities - plasmas -pulsars: general - pulsars: individual: J2144-3933
\end{keywords}

\section{Introduction}

A half century has elapsed since the discovery of pulsars by Hewish et al (1968) and their prompt recognition as rotating magnetized neutron stars (Pacini 1967, Gold 1968).  The conclusions of some early seminal papers (Goldreich \& Julian 1969, 
Radhakrishnan \& Cooke 1969) remain essentially good although the former is subject to the later development of the force-free approximation and of the equatorial current sheet external to the light cylinder by Contopoulos, Kazanas \& Fendt (1999).  But attempts to provide a physical basis for polar-cap generation of coherent radio emission have resulted in little progress apart from the outstanding work of Beskin (1990) and Muslimov \& Tsygan (1992) on the role of the Lense-Thirring effect in particle acceleration.

In a previous paper (Jones 2018) we have expressed the view that this remarkable lack of progress is a direct consequence of assumptions about electron-positron pair production above polar caps for which there is no physical basis. More generally,
recent work on the equatorial current sheet has emphasized that the force-free model of the magnetosphere close to the light cylinder does not require electron-positron production anywhere in the magnetosphere to maintain a stable net charge on the star
(see Contopoulos 2016).

It is a commonplace remark that radio emissions considered to be of polar-cap origin are broadly similar for the majority of both normal and millisecond pulsars (MSP).  The assumption of electron-positron pair creation needs to be justified in the many cases for which it is clearly implausible such as most of the MSP and particularly in the case of PSR J2144-3933 whose first observers (Young, Manchester \& Johnston 1999) raised this specific question in their paper.  This is true for both aligned ($\Omega\cdot{\bf B} > 0$) and counter-aligned ($\Omega\cdot{\bf B} < 0$) neutron stars, where $\Omega$ is the rotation spin and ${\bf B}$ the polar-cap magnetic flux density. Aligned pulsars may be thought not to have the degrees of freedom at the polar cap necessary to account for the phenomena, nulls, mode-changing, and sub-pulse drift that are observed unless it is supposed that the physics of a compact high-field high energy-density environment is controlled by the state of light-cylinder plasma. Counter-aligned pulsars are subject to well-established nuclear processes that have simply been ignored, but are crucial.  We refer to proton production in electromagnetic showers produced by a reverse-electron flux.  Electron-positron production, if present, must be accompanied by a flux of protons that has been ignored.  Only in rare cases can pairs be produced and then only by inverse Compton scattering.  In fact, the principal source of reverse-electron energy is the photo-ionization of ions accelerated through the blackbody radiation field of the neutron-star surface.    
Our belief is that the ion-proton system is the source of coherent emission at frequencies of the order of 1 GHz in virtually all pulsars, normal and millisecond.

 The present paper describes a model polar cap based on these physical processes which have been examined in a number of previous papers (see Jones 2016a for a summary).  The construction of a model polar cap based on precisely defined physical processes was not attempted in those papers and its lack is the motivation for this work.  It shows that a model source of plasma can be constructed which can have the pulse-to pulse variability seen in the observed emission and should be capable of developing strong turbulence. It also demonstrated the parameters on which this variability depends.  With increasing mean acceleration potential above the polar cap, nulls appear and in the limit the state of the polar cap approaches that characteristic of Rotating Radio Transients (RRAT).
 
 An important feature of the model is its consistency with the observed slowly and smoothly varying longitude-dependence of the circular polarization that is found in many radio pulsars (we refer in particular to the high-resolution measurements of Karastergiou \& Johnston 2006; see Jones 2016b).  The passage of radiation through the magnetosphere post-emission (after the surface of last scattering) depends on the degree of birefringence, that is, the difference between the path integrals of the O- and E-mode refractive indices.  Because the difference between the refractive indices  is a function of plasma particle Lorentz factor, which must vary considerably with observational longitude, the form of the circular polarization implies that the birefringence is very small compared with that for a high-density electron-positron plasma.  The point is not that circular polarization is present but that its longitude variation is slow and smooth at high resolution.  Small birefringence is consistent with a plasma of nucleon rest masses.  It is also true that the radio emission cyclotron absorption problem, widely discussed in previous literature on electron-positron plasmas, is no longer present.
 
 The model is described fairly completely in Section 2 and its qualities as a source of coherent emission are examined in Sections 3 and 4.  Its stability, also its plausibility and
adaptability in relation to neutron-star parameters that are not all well known are considered in Section 5.  A Fortran file of model calculation algorithms can obtained by e-mail from the author.

We have to emphasize that the model is not concerned with coherent radiation whose source is in the region of the light cylinder, as in the Crab pulsar main pulse and inter-pulse. Recent developments in this subject, based on the concept of magnetic re-connection, have been described by Lyubarsky (2019) and Philippov et al (2019).

\section{The model}

An account of the model starts with  the polar cap of radius $u_{0}$.  It is assumed (probably incorrectly) to be circular and encloses those magnetic flux lines, termed open, that pass through the light cylinder but not through the null (zero charge-density) surface.  We assume a zero-potential boundary condition on the surface enclosing this part of the magnetosphere in the corotating frame and space-charge limited flow at the neutron-star surface for protons and ions.
The model acceleration field ${\bf E}_{\parallel}$ inside the open sector is time-varying but not so rapidly that retardation becomes significant. Solutions of Poisson's equation are required in huge numbers, but fortunately, the geometrical shape of the open sector above the polar cap allows a simple and fast approximation, adequate for the model.  The radius of the open sector for a dipole field is,
\begin{eqnarray}
u_{0}(\eta) = \left(\frac{2\pi R^{3}\eta^{3}}{cPf(1)}\right)^{1/2},
\end{eqnarray}
(Harding \& Muslimov 2001), where $\eta$ is the radial polar coordinate in units of the neutron-star radius $R$.  For a $1.4$ ${\rm M}_{\odot}$ mass and radius $R = 1.2\times 10^{6}$ cm, the constant $f(1)= 1.368$.  At altitudes below $\eta \approx 4$, the change $\delta u_{0}$ in radius occurring within an altitude difference of the order of $u_{0}$ is very small compared with $u_{0}$. Thus the system is a long narrow tube which can be represented locally by an infinite right cylinder of constant radius $u_{0}$.  (This approximation is less good in the case of MSP:  for this reason, model computations have not been made for these pulsars.) The potential Green function, under Dirichlet boundary conditions, for a line source in such an infinite cylinder and in cylindrical polar coordinates $(u,\phi)$ is
\begin{eqnarray}
G({\bf u}, {\bf u}^{\prime}) = \ln\left(\frac{u_{0}^{4}+u^{2}{u^{\prime}}^{2}-
2u_{0}^{2}uu^{\prime}\cos(\phi -\phi^{\prime})}{u_{0}^{2}u^{2}
 + u_{0}^{2}{u^{\prime}}^{2} - 2u_{0}^{2}uu^{\prime}\cos(\phi - \phi^{\prime})}\right).
\end{eqnarray}
Poisson's equation is also easily solved at altitudes $\ll u_{0}$ where the problem is one-dimensional (see Mestel et al 1985, Beloborodov 2008).  The intermediate region, altitudes of the order of $u_{0}$, is where the principal interaction of ions is with the polar-cap blackbody field, temperature $T_{pc}$. The electric field here can be estimated only by interpolation but is not important except in cases where the neutron-star whole-surface temperature $T_{s}$ is negligible.

The acceleration potential at altitudes $\gg u_{0}$ for a charge density $\sigma$ uniform over the open sector is, in the cylinder approximation,
\begin{eqnarray}
\Phi(u,\eta) = \pi(u_{0}^{2} - u^{2})(\sigma - \sigma_{GJ}),
\end{eqnarray}
in which the Goldreich-Julian charge density is
\begin{eqnarray}
\sigma_{GJ} = - \frac{B}{cP\eta^{3}}
\left(1 - \frac{\kappa}{\eta^{3}}\right)\cos\psi,
\end{eqnarray}
where $\psi$ is the angle subtended by the polar-cap magnetic flux density ${\bf B}$ and the rotation spin $\Omega$.  The dimensionless Lense-Thirring factor is $\kappa = 0.15$ (Harding \& Muslimov (2001).  We set $\cos\psi = -1$ and with the charge density,
\begin{eqnarray}
\sigma = - \frac{B}{cP\eta^{3}}\left(1 - \kappa\right)
\end{eqnarray}
as would be the case for acceleration of a flux of protons only, the potential is an appropriate reference potential and $\sigma$ a reference charge density.

The potential in the presence of an ion component is much reduced through screening by the flux of photoelectrons produced by the interaction of the accelerating ions with the whole-surface blackbody field.  Photo-ionization has low probability for $\eta > 4$ owing to the nature of the Lorentz transformation from the corotating frame to the instantaneous rest frame of the ion.  The reverse flux of electrons changes the charge density of the ion component but the proton component is unaffected.  At a small altitude $h$ of the order of $u_{0}$ the charge density prior to interaction with the blackbody field is,
\begin{eqnarray}
\sigma(h) = {\rm e}N_{z}(2Z_{h} - Z_{\infty} +K) = \sigma_{GJ}(h),
\end{eqnarray}
where $N_{z}$ is the ion number density, $Z_{h}$ is the prior and $Z_{\infty}$ the final ion charge defined as at $\eta = 4$.  The number of protons is $K$ per ion.  (For small distances above the polar cap it is convenient to define the altitude $h = (\eta - 1)R$.) The final charge density following photo-ionization is,
\begin{eqnarray}
\sigma_{\infty} = {\rm e}N_{z}(Z_{\infty} + K).
\end{eqnarray}
As the ion accelerates, the charge density increases smoothly from equation (6) to
equation (7) closely following the Lense-Thirring change in $\sigma_{GJ}$ and screening the acceleration potential to values perhaps an order of magnitude or more  smaller than the reference field.

Reverse electrons follow closely the ion path back to the neutron-star surface and create electron-photon showers in the compact atmosphere whose scale height is of the order of $10^{-1}$ cm.  Proton formation by decay of the giant dipole state occurs preferentially at the shower depth, about $10$ radiation lengths, where the track length of $15 - 20$ MeV photons is a maximum. The steps in the rate calculation are as follows.  The total photon path length per unit interval of photon momentum in a shower is well known and is, for low energy photons, a linear function of initial electron energy.  The total cross-section for nuclear photo-absorption, integrated over energy in the region of the giant dipole state, is also well known and for low nuclear charge the partial widths for decay by neutron or proton emission are approximately equal. The rate of proton formation with energies of the order of $2-3$ MeV is known: the rates for further nuclear reactions are negligible for protons and they are rapidly thermalized.
Protons cannot be in static thermal equilibrium in a predominantly electron-ion atmosphere:  they are subject to the small $E_{\parallel}$ which maintains the electron-ion equilibrium in the gravitational field and therefore move to the top forming a further atmosphere above the ions.  The important point is that as a consequence, they are accelerated  preferentially compared with the ions.  At any instant, ions are accelerated only if there are insufficient protons to satisfy the Goldreich-Julian charge density given by equation (6). It has to be emphasized that the process described here is quite accurately local.  The number of protons $W_{p}$ created per unit electron energy is a function of $B$ and of the nuclear charge $Z_{0}$ in the very outer layers of the neutron-star crust and the extent to which it is reduced by giant dipole state formation. But we have assumed a constant value, $W_{p} = 0.2$ GeV$^{-1}$.  We refer to Jones (2010) for a more complete account of proton formation.

In principle, the polar-cap area should be treated as a continuum but division into a finite number of discrete equal-area cylindrical elements has been adopted.  The central element is coaxial with the magnetic axis.  The remaining elements form a circular bundle being placed in $N$ coaxial rings of $6, 12, ....$ elements, the total number being $N_{c} =1 + 3N +3N^{2}$. For geometrical positioning only, each element is notionally of radius
$u_{0}/(2N + 1)$ and their angular positions are non-overlapping but chosen randomly.
However, for electrostatic calculation, each element is of radius $u_{0}/N_{c}^{1/2}$. The notional overlap that this entails is permissible because the interaction between two elements is approximated as being that between two line sources. The electrostatic potential on the axis of each element is required at an altitude $\eta \approx 4$ where the ion charges $Z_{\infty}$ will have stabilized.  In the general state, elements have different excess charge densities $2{\rm e}N_{z}(Z_{\infty} - Z_{h})$ above the reference value. The potential at element $i$ is then given by,
\begin{eqnarray}
\Phi({\bf u}_{i}) = \Phi_{ref} + 2{\rm e}\int_{i}d{\bf u} G({\bf u}_{i},{\bf u})
N_{z}^{i}(Z^{i}_{\infty} - Z_{h})         \nonumber \\
 + 2{\rm e}\sum_{j\neq i} \int d{\bf u} 
G({\bf u}_{i}, {\bf u})N^{j}_{z}(Z_{\infty}^{j} - Z_{h})\delta({\bf u} - {\bf u}_{j}).
\end{eqnarray}
In the case of element $j = i$, the integration is over the area of that element, otherwise, elements $j\neq i$ are treated as line sources positioned as we have described above, a fast and simple procedure adequate for the present model and consistent with our treatment of element areas.  But the $\Phi_{i}$ must be compatible with the individual $Z^{i}_{\infty}$.  

The relation between these quantities has been obtained by calculating the photoelectric transition rates for ions accelerated from the polar cap and interacting with the whole-surface blackbody field but subject to the reference potential $\Phi_{ref}$ defined by equations (3)-(5) with a cut-off $\Phi_{c}$. At a given $\eta$, the transition rate for photo-ionization by the blackbody field of the visible neutron-star surface is obtained
leading by integration to average values for the number of reverse electrons $Z_{\infty} - Z_{h}$ per ion and their total energy $\epsilon_{s}$ at the neutron-star surface.
  
Calculation neglects red shifts and assumes rectilinear rather than general-relativistic photon paths. Photo-ionization cross-sections are large immediately above threshold though not well known and the principal factor is the acceleration of the ion to the extent that blackbody photons reach the threshold in the ion rest frame.  We refer to Jones (2012) for more complete details.  Thus functions $Z_{\infty}(\Phi_{c})$ and $\epsilon_{s}(\Phi_{c})$ are known, where $\epsilon_{s}$ is the energy flux of reverse electrons per ion at the polar cap surface resulting from interaction with the whole-surface field $T_{s}$. This is not an observer-frame temperature but is better approximated by that of the local proper frame. Within each time-step, self-consistent solutions of $Z^{i}_{\infty}(\Phi_{i})$ and of equation (8) are obtained by a simple relaxation procedure involving a large number of successive small independent adjustments of the $Z^{i}_{\infty}$.

It is convenient to introduce the variable,
\begin{eqnarray}
\alpha_{p} = {\rm e}\frac{KN_{z}}{\sigma_{GJ}(h)}
\end{eqnarray}
representing the proton fraction of the charge density at any given time.  Then,
\begin{eqnarray}
\frac{{\rm e}N_{z}}{\sigma_{GJ}(h)} = \frac{1 - \alpha_{p}}{2Z_{h} - Z_{\infty}}.
\end{eqnarray}
The number of protons created per ion within a time-step is then
$\tilde{K} = W_{p}\epsilon$, with $\epsilon = \epsilon_{h} + \epsilon_{s}$, and where the constant $W_{p} = 0.2$ GeV$^{-1}$ (see Jones 2010).  Here $\epsilon_{h}$ is the energy flux of reverse electrons from interaction with the polar-cap blackbody field $T_{pc}$ at low altitudes of the order of $u_{0}$.  In most normal pulsars, $\epsilon_{h} \ll \epsilon_{s}$ but in exceptional instances in which $T_{s}$ is very low this term becomes significant.  It is further considered in Section 4 in relation to some particular pulsars.  The quantity $\tilde{K}$ is distinct from $K$ because these protons are created in the electron-photon shower and reach the surface only in later steps. The mean time for proton transport from the shower maximum to the top of the atmosphere is $\tau_{p}$.  Thus protons produced within a given integration step are accelerated only during later steps. The proton charge density reaching the surface notionally at the start of a step $\Delta$ is,
\begin{eqnarray}
\tilde{\alpha}_{p} = W_{p}\sum_{\Delta^{\prime}}\left(\frac{\epsilon(1 - \alpha_{p})
}{2Z_{h} - Z_{\infty}} \right)_{\Delta^{\prime}}f_{p}(\Delta^{\prime}),
\end{eqnarray}
in units of $\sigma_{GJ}(h)$, where the sum is over all previous steps $\Delta^{\prime}$.  The normalized weight function $f_{p}$ represents the distribution of times needed for  protons to move from the shower maximum to the surface under the influence of the small $E_{\parallel}$.  This is a Gaussian centred at $\tau_{p}$ with half-width $\Delta$: owing to the presence of $E_{\parallel}$ in the atmosphere it is a compromise between a $\delta$-function and a zero-field diffusion function.  (The magnitude of $\tau_{p}$ is of the order of $1$ s but is not well known, see Jones 2011, 2013).
If $\tilde{\alpha}_{p} < 1$, it is the value of $\alpha _{p}$ for the time-step which is commencing.  But if $\tilde{\alpha}_{p} > 1$, no ions can be accelerated in that time-step within which $\tilde{\alpha}_{p}$ is reduced to $\tilde{\alpha}_{p} - 1$.

The time-steps have been chosen to be of length $\Delta = 0.2\tau_{p}$.  A step commences with the known state of the proton atmosphere for each element found from its previous history as described above.  This determines whether a mixture of protons and ions or of protons only is accelerated, and the step ends with the computation of proton creation (if any) in that step and re-setting the state of the proton atmosphere.  The output consists of the sequence of values $Z^{i}_{\infty}$ and $\Phi_{i}$ over the polar cap.

To be a source of plasma turbulence, an element must have an ion-proton composition with ion and proton Lorentz factors below critical values such that the Langmuir-mode growth rate is large enough for the development of non-linearity.  The ion and proton velocity distributions are essentially $\delta-$functions and therefore the system is ideally unstable against growth of a longitudinal or quasi-longitudinal mode.  That the former does not couple with the radiation field is of no consequence if strong turbulence evolves.  We have to assume that the growth-rate of the mode which has been obtained only in the linear approximation continues to non-linearity and the development of strong turbulence whose decay is the source of the coherent emission.  The distribution of radiation angle $\chi$ at the emission altitude with respect to the element axis is determined by the Lorentz factors.  We assume isotropy in the turbulence rest frame so that the intensity generated by an element is $\propto
\gamma^{2}/(1 + \gamma^{2}\chi^{2})^{2}$. The intensity at a point on the arc of traverse is defined here as the sum over all elements of this factor.  Thus actual generation of coherent emission itself is not calculated here, merely the conditions necessary for it to occur.  The typical frequency of such radiation must be of the order of the particle plasma frequency concerned.  Protons and ions have plasma frequencies at number densities for $\eta < 10$ that are consistent with radiation below $1$ GHz.  (It is worth noting that for a high-density electron-positron plasma the frequencies are some orders of magnitude too high unless the region of emission is at large values of $\eta$ that would be beyond the light cylinder in the MSP.) The intensity generated by the whole polar cap is then calculated as a function of observational longitude on a (straight-line) arc of traverse crossing the polar cap. The variability inherent in the state of the polar cap has a characteristic time less than or of the order of $\tau_{p}$.  This is not related to the rotation period $P$ but depends on the condition of the electron-ion atmosphere.  The rotation period enters only the calculation of $\Phi_{ref}$ and the radius of the polar cap $u_{0}$ which affects the  estimate of $\epsilon_{h}$.  The integration step-length $\Delta$ is related to $\tau_{p}$: it is coincidental that this is of the same order as a typical normal pulsar period. Null states reflect either the absence of emission from all elements  in any given step, or emission not exceeding $10^{-2}$ of the typical mean intensity.

\section{Model polar caps}
In order to display model calculation results, the polar cap is mapped on to a circular surface at altitude $\eta$ and of radius $\theta_{0} = 3u_{0}(\eta)/2\eta R$, thus showing the angles of tangents to open magnetosphere flux lines.  Observations on normal pulsars (Hassall et al 2012) and on millisecond pulsars (MSP; Kramer et al 1999) indicate low emission altitudes and the value $\eta = 5$ has been adopted here.  The (straight-line) arc of traverse through the line of sight has a minimum angle $\theta_{0}/4$ with the magnetic axis.

Diagrams showing limited numbers of individual emission profiles do not convey much information. Step-to-step variability is the significant information about the emission profile and is given here by the longitude-resolved intensity modulation index, as defined by Jenet \& Gil (2003), calculated at points $\pm 0.4\theta_{0}$ and $\pm 1.2\theta_{0}$ with respect to the profile centre, chosen to sample core and conal sectors. This is the normalized root-mean-square deviation evaluated at the specified longitude.  Modulation indices here reflect only variability in the precursor state of emission.  Further stochastic components will also arise from turbulence formation and decay and naturally contribute to the observed variability.

Coherent emission requires plasma with both ion and proton components having only moderate Lorentz factors and hence a Langmuir-mode growth rate capable of generating non-linearity and turbulence.  A critical value $\gamma_{c} = 100$ for the proton component has been adopted by reference to previous work (Jones 2016a) but is to some extent  arbitrary because the conditions from which growth initiates are unknown.  Specifically, it is difficult to exclude the possibility not considered here of oscillatory behaviour in the low-altitude region $(h \sim u_{0})$ above the neutron-star surface.  Time-dependent currents and charge densities at frequencies of the order of the local plasma frequency are likely to assist the growth of the Langmuir mode to the limit of non-linearity and turbulence generation that we assume.

The amplitude gain factor $\exp(\Lambda)$ for the mode is given by
\begin{eqnarray}
\Lambda = Im(s)\left(\frac{4\pi(1 - \alpha_{p})Z_{\infty}{\rm e}R^{2}B}
{PAm_{p}c^{3}}\right)^{1/2}\\    \nonumber
\int^{\infty}_{1}d\eta
\left(\frac{1}{\eta\gamma_{z}(\eta)}\right)^{3/2},
\end{eqnarray}
in terms of the ion Lorentz factor at altitude $\eta$ and the slowly-varying dimensionless quantity
$Im(s)\approx 0.2$ for a typical pulsar; see Jones (2014).  A value $\Lambda = 30$
has been assumed previously to be adequate for producing growth to turbulence, but much lower values would suffice if low-altitude oscillatory behaviour exists.  A computed value of $\Lambda$ would require knowledge of the acceleration potential at all altitudes.  This is beyond the scope of the present model which assumes a $\Phi_{ref}$ subject to a constant cut-off.   

An element in which $\gamma_{p} > \gamma_{c}$ or with no ion component$(\alpha_{p} =1)$ in a step does not radiate.  Both the total number of null steps  generated by the model and the lengths of individual nulls have been recorded.

Values of the surface nuclear charge $Z$ are not known.  At depths below the surface large compared with electron-photon shower depths we assume the canonical value $Z_{0} = 26$.  But photo-disintegration implies that the mean value at the top of the electron-ion atmosphere must be approximately related by $\langle Z \rangle = Z_{0} - \langle \tilde{K} \rangle$ though with some modification possible through $\beta$- processes.   The values adopted here, $\langle Z \rangle = 10$ and $20$ are consistent with a range of reasonable values for $\langle \tilde{K} \rangle$.  The condition of matter immediately below the surface is uncertain: a liquid state is not impossible.
Furthermore, the $\tilde{K}$ protons are produced at a depth of $\sim 10$ radiation lengths and this can lead to instabilities in the value of $\langle Z \rangle$ on time-scales of the order of the ablation time for acceleration of one radiation length of ions
\begin{eqnarray}
\tau_{rl} = 2.1\times 10^{5}\left(\frac{P}{\langle Z \rangle B_{12} \ln (
12\langle Z \rangle^{1/2}B_{12}^{-1/2})}\right) \hspace{3mm} s,
\end{eqnarray}
(Jones 2011).  For $Z = 10$ and $20$, we adopt ion charges $Z_{h} = 8$ and $17$  respectively for $B_{12} =0.1$, $8$ and $14$ for $B_{12} = 1.0$, and $6$ and $10$ for $B_{12} = 10.0$, to be inserted in equation (6). These degrees of ionization are  derived from the free-electron chemical potential, at the density estimated for the top of the atmosphere and in a field of $10^{12}$ G, which we find to be $\approx -26k_{B}T$ (see Jones 2012).

Results are given for values of $B$ and $P$ that span the observed pulsar population with the exception of MSP to which we make reference later.  Values of $T_{s}$ large enough to be of importance are in the interval $2 - 4\times 10^{5}$ K in the proper frame of the neutron-star surface.  Neutron stars with lower $T_{s}$ rely on $T_{pc}$ or on other processes to produce a reverse flux of electrons and are considered later in Section 4.  PSR J2144-3933 is known to be a member of this class with an observed whole-surface temperature of $< 4.2\times 10^{4}$ K (Guillot et al 2019), found by Hubble Space Telescope observation of the Rayleigh tail of the blackbody spectrum.  Such observations have lead to estimates $5\times 10^{4} - 2.6\times 10^{5}$ K for the MSP J2124-3358 (Rangelov et al 2017) and $1.3 - 2.5\times 10^{5}$ K for B0950+08 (Pavlov et al 2017) which are not obviously inconsistent with the $T_{s}$ values adopted here for calculation.

\begin{table*}
\caption{For sets of $10^{3}$ steps at magnetic fields $B$, period $P$ and whole-surface temperature $T_{s}$, the number of nulls in radio emission and their mean lengths (steps) are given. The phase-resolved intensity modulation indices are, in order, from $-1.2\theta_{0}$ to $1.2\theta_{0}$, where $\theta_{0}$ is the angle subtended by the tangents to the open-closed surface at an altitude $\eta = 5$. The line of sight is off-set from the magnetic axis by an angle $\theta_{0}/4$. The upper values are for surface atomic number $Z = 10$ and the lower for $Z = 20$.}

\begin{tabular}{@{}lrrrrrrrr@{}}
\hline
 $B$ & $P$  & $T_{s}$ & nulls & length & $m_{1}$ & $m_{2}$ & $m_{3}$ & $m_{4}$  \\
\hline
 $10^{12}$ G &  s & $10^{5}$ K &   &   &   &  &  &  \\
\hline
  10.0 & 1.0  &   2  & -  &  - & -  & - & - & -  \\
        &     & 3  & -  & -   & -& - & - & - \\
        &     & 4  & 201 & 2.8 & 2.95 & 6.79 & 6.67 & 2.27  \\
    & 3.0 &  2 & - & - & - & - & - & - \\
        &     &  3 & 162 & 4.3 & 4.03 & 3.44 & 3.50 & 3.79  \\
        &     &  4 &  5  & 1.8 & 0.24 & 0.47 & 0.48 & 0.25  \\
    & 5.0 &  2 &  0  & 0 & 0.01 & 0.00 & 0.00 & 0.01  \\
        &     &  3 &  0  & 0 & 0.03 & 0.02 & 0.02 & 0.03  \\
        &     &  4 &  0  & 0 & 0.16 & 0.29 & 0.30 & 0.12  \\
   1.0  & 0.5 & 2  &  -  & - & -   &   -  &  -   &  -   \\
        &     & 3  & 186 & 4.0 & 2.92 & 8.34 & 7.92 & 2.72  \\
        &     & 4  &  45 & 1.1 & 1.10 & 1.44 & 1.46 & 1.08  \\
        & 1.0 & 2  & 15 & 53.0 & 11.10 & 10.44 & 11.54 & 11.81  \\
        &     & 3  &  11& 1.2 & 0.28 & 0.83 & 0.83 & 0.30  \\
        &     & 4  &   3 & 1.3 & 0.26 & 0.63 & 0.63 & 0.27  \\
        & 2.0  & 2  &  0  & 0 &  0.01 & 0.00 & 0.00 & 0.01  \\
        &      & 3  &  0  & 0 & 0.11 & 0.31 &  0.31 & 0.10  \\
        &      & 4  &  0  & 0 & 0.08 & 0.25 & 0.24 & 0.09  \\
   0.1 & 0.5   & 2  &  0  & 0 & 0.07 & 0.47 & 0.48 & 0.08  \\
        &      & 3  &  0  & 0 & 0.02 & 0.05 & 0.04 & 0.02  \\
        &      & 4  &  0  & 0 & 0.01 & 0.03 & 0.03 & 0.01  \\
 \hline
   10.0 & 1.0  & 2  &  -  &  -  &  -  &  -  &  -  \\
        &      & 3  &  -  &  -  &  -  &  -  &  -  \\
        &      & 4  & 145 & 5.5 & 4.35 & 9.98 & 9.22 & 4.40  \\
        & 3.0  & 2  &  -  &  -  &  -  &  -  &  - & -    \\
        &      & 3  &  29 & 30.2 & 12.01 & 11.74 & 10.20 & 10.64  \\
        &      & 4  &   3 & 1.3 & 0.28 & 0.55 & 0.55 & 0.28  \\
        & 5.0  & 2  &  0  & 0 & 0.01 & 0.00 & 0.00 & 0.01  \\
        &      & 3  &  0  & 0 & 0.03 & 0.02 & 0.02 & 0.03  \\
        &      & 4  &  0  & 0 & 0.11 & 0.32 & 0.33 & 0.11  \\
   1.0  & 0.5  & 2  &  -  &  -  &  -  &  -  &  -& -  \\
        &      & 3  & 116 & 1.2 & 1.33 & 2.45 & 2.34 & 1.34  \\
        &      & 4  &   4 & 1.5 & 0.37 & 0.57 & 0.54 & 0.40  \\
        & 1.0  & 2  & 136 & 5.8 & 4.48 & 3.84 & 3.48 & 4.04  \\
        &      & 3  &   0 & 0 & 0.12 & 0.58 & 0.59 & 0.13  \\
        &      & 4  &   0 & 0 & 0.12 & 0.34 & 0.34 & 0.12  \\
        & 2.0  &  2 &  0  & 0 & 0.02  &  0.09  &  0.09  & 0.02  \\
        &      & 3  &  0  & 0 & 0.12 & 0.36 & 0.37 & 0.12  \\
        &      & 4  &  0  & 0 & 0.04 & 0.09 & 0.08 & 0.04  \\
   0.1  & 0.5  & 2  &  0  & 0 & 0.03 & 0.13 & 0.13 & 0.03  \\
        &      & 3  &  0  & 0 & 0.05 & 0.58 & 0.58 & 0.05  \\
        &      & 4  &  0  & 0 & 0.03 & 0.41 & 0.42 & 0.03  \\  
\hline
\end{tabular}
\end{table*}

The $\theta_{0}/4$ offset of the arc of traverse from the magnetic axis exposes the presence of both conal and core-profile components.  Suppression of the core component occurs owing to the higher values of the potential in that region which produce the conditions $\gamma_{p} > \gamma_{c}$ or $\alpha_{p} =1$ more frequently. The conal components then appear as a double-peaked profile.

The modulation indices, numbers of nulls and their mean lengths are given in Table 1 for sets of $10^{3}$ steps with nuclear charges $Z = 10$ and $20$.  A blank entry represents a complete sequence of $10^{3}$ null steps.  In such a case, a re-run with an increased value of $\gamma_{c}$ produces other than null steps but which cannot be radio emission precursors.

The reference potential given by equations (3)-(5) is $\Phi_{ref} \propto BP^{-2}$ and the extent to which it is screened by the reverse flux of electrons determines the average acceleration potential $\langle \Phi_{i} \rangle$ in each element that has $\alpha_{p} < 1$. (Proton-only elements are of no concern in any step.) It is found that at constant $B$ and $T_{s}$, $\langle \Phi_{i} \rangle \propto \Phi_{ref}^{1/2}$ approximately.  Thus $\langle \Phi_{i} \rangle \propto \Phi_{ref}^{1/2}T_{s}^{-1}$ over limited intervals of $P$ and $T_{s}$ owing to photo-ionization cross-sections being large near thresholds so that the ion Lorentz factor at ionization is $\propto T_{s}^{-1}$ approximately.

Table 1 entries are consistent, for both $Z = 10$ and $20$, with these two variables controlling the frequency and average length of nulls and the magnitude of conal and core modulation indices.  Small $\langle \Phi_{i} \rangle$ leads to few null steps and relatively little step-to-step profile variation. This can be seen by reference, for example, to the $Z =10$, $B = 10^{12}$ G, $T_{s} = 3\times 10^{5}$ K entries for different values of $P$, and therefore of $\Phi_{ref}$.  There are $186$ nulls of average length $4.0$ and large core modulation indices at $P = 0.5$ s, but only $11$ nulls of average length $1.2$ at $P = 1.0$ s with much reduced modulation indices.  At $P = 2.0$ s, modulation indices are small with no nulls.  Varying $T_{s}$ at constant $P$ further demonstrates the dependence of nulls on $\langle \Phi_{i} \rangle$. 

 A large fraction of normal pulsars appear to exhibit no nulls and these have small $\Phi_{ref}$. Their precursor modulation indices are small and it must be presumed that their observed variability derives almost completely from the decay of turbulence or from passage through the inter-stellar medium. 
The model is intended to give a general physics-based understanding of the emission process and its evolution and should not be compared with data for any specific pulsar:  the important parameters $Z$ and $T_{s}$ are not known in any specific case.
Values of $T_{s} > 4\times 10^{5}$ K have not been investigated, but we can observe that small $\langle\Phi_{i} \rangle$ and hence small ion and proton Lorentz factors imply emission in the observer frame at significant angles to local flux lines and some profile widening.

The influence of $\langle \Phi_{i} \rangle$ on null formation has been examined in more detail firstly for the specific case $Z = 20$, $B_{12} = 10$ and $T_{s} = 3\times 10^{5}$ K.  Table 2 gives the null frequency and average lengths for a compact interval of $P$ centred on $P = 3.0$ s for sets of $2000$ steps.  Null frequency is large at $P =3.2$ s but average length is short.  As $\langle \Phi_{i} \rangle$ increases, the frequency falls with increasing average length until ultimately radio emission ceases.  During this phase, the null length distribution, shown in Fig. 1, changes to one that is immediately suggestive of the Rotating Radio Transients (RRAT).  The values of $B$ and $P$ chosen for Table 2 are broadly an average of those listed in the RRAT catalogue of Keane et al (2011).  These results are consistent with the RRAT being the final state of large $B$ pulsars that have passed through an epoch of nulls with increasing $\langle \Phi_{i} \rangle$.

\begin{table}
\caption{Here the information is at $Z = 20$, $B = 10^{13}$ G, and $T_{s} = 3.0\times 10 ^{5}$ K for a compact set of periods $P = 2.8 - 3.2 $ s in sets of $2\times10^{3}$ steps. Nulls, mean lengths and modulation indices are given as in Table 1. }
\begin{tabular}{@{}rrrrrrr@{}}
\hline
    $P$      &  Nulls  & length & $m_{1}$ & $m_{2}$ &$m_{3}$& $m_{4}$\\
\hline
   3.2 &    380 & 3.0 & 2.8  & 2.3  & 2.4  & 3.0   \\
        3.1 &    278 & 5.6 & 4.4 & 3.7 & 3.9 & 4.6  \\
         3.0 &    115 & 16.0 & 8.1 & 7.3 &8.6 & 9.7  \\
         2.9 &     25 & 49.5 & 16.3 & 15.6 &14.9 & 15.0 \\
         2.8 &   - & - & - & - & - & -  \\
\hline
\end{tabular}
\end{table}

For a specific neutron star, it is likely that $\langle \Phi_{i} \rangle$ is largely determined by the natural decrease in $T_{s}$ with age.  Consequently, evolution starting from a state of relatively small $\langle \Phi_{i} \rangle$ and stable coherent radio emission proceeds firstly to a state of nulls with increasing null frequency followed, ultimately, by decreasing frequency as the average length increases.  This is demonstrated in Table 3 which is for $Z = 20$, $B = 10^{13}$ G,
$P = 3.0$ s and a temperature interval from $3.2 - 2.8 \times 10^{5}$ K in sets of
$2\times 10^{3}$ steps.  Nulls, mean lengths and modulation indices are given as in Table 2.
A cut-off defined principally by the whole-surface temperature $T_{s}$ has been described previously (see Jones 2015; Fig.1) but is now supported by model calculation.

\begin{table}
\caption{This is for $Z = 20$, $B = 10^{13}$ G and $P = 3.0$ s and as a function of
$T_{s}$ (in units of $10^{5}$ K) and in sets of $2\times 10^{3}$ steps}

\begin{tabular}{@{}rrrrrrr@{}}
\hline
  $T_{s}$  &  Nulls  & length & $m_{1}$ & $m_{2}$ & $m_{3} $ & $m_{4}$  \\
  
\hline
	3.2  &  38  &  1.2  &  1.0  &  0.8  &  0.8  &  1.0  \\
	3.1  &  206  &    1.2  &  1.5  &  1.2  &  1.2  &  1.5  \\
	3.0  &  415  &  2.0  &  2.4  &  1.9  &  2.0  &  2.5  \\
	2.9  &  246  &  6.5  &  5.0  &  4.5  &  4.8  &  5.6  \\
	2.8  &   32  &  61.0  &  16.0  &  16.0 & 14.6  &  15.1  \\
	
\hline
\end{tabular}
\end{table}

The cut-off proton Lorentz factor for Tables 1 - 3 is $\gamma_{c} = 100$.  Table 4
gives data for a more archetypal period and magnetic flux density; $P = 1.0$ s,
$B = 10^{12}$ G, in sets of $2\times 10^{3}$ steps subject to a reduced cut-off $\gamma_{c} = 50$.  Here, the effect of cooling is less rapid than at $10^{13}$ G.
The data are given as in Tables 2 and 3 and the distributions of nulls are broadly
similar to that in Fig. 1.

The numbers of nulls in any case are not large and any seeming discrepancy between different Tables in their frequency and mean length is not obviously inconsistent with the use of different random number sets in defining the initial conditions of the calculation, specifically, the initial value of $\alpha_{p}$ for each cell.  That the output after many steps is very dependent on the initial conditions is not surprising: the system is causal but non-linear.

\begin{table}
\caption{These data are for $Z = 20$, $B = 10^{12}$ G, and $P = 1.0$ s in sets of $2\times 10^{3}$ steps.  Here the cut-off is at the lower value $\gamma_{c} =50$.}
\begin{tabular}{@{}rrrrrrr@{}}
\hline
  $T_{s}$ & Nulls & length & $m_{1}$ & $m_{2}$ & $m_{3}$ & $m_{4}$   \\
\hline
   3.0  &    2  &  1.0  &  0.1  &  0.3  &  0.2 &  0.1  \\
   2.8  &   2  &  1.5  &  0.1  &  0.2  &  0.2 & 0.1  \\
   2.6  &  4  &  1.5  &  0.3  &  0.3  &  0.3 & 0.2  \\
   2.4  &  8  &  1.7  &  0.6  &  0.7  &  0.7 & 0.7  \\
   2.2  &  172  &  1.1  &  1.3  &  1.5  &  1.5  &  1.3  \\
   2.0  &  554  &  1.7  &  2.6  &  3.8  &  4.0  &  2.7  \\
   1.8  &  276  &  6.1  &  6.0  &  9.3  &  9.8 & 6.4  \\
   1.7  &  163  &  11.1  &  9.0  &  12.8  &  12.1  &  8.1  \\
   1.6  &  66  &  28.4  &  12.9  &  16.7  &  15.5  &  10.8  \\
   1.5  &  18  &  75.5  &  16.1  &  18.4  &  18.7  &  17.0  \\
   1.4  &  -  &  -  &  -  &  -  &  -  &  -  \\
\hline
\end{tabular}
\end{table}

\begin{figure*}
\includegraphics[trim=10mm 85mm 20mm 80mm, clip, width=168mm]{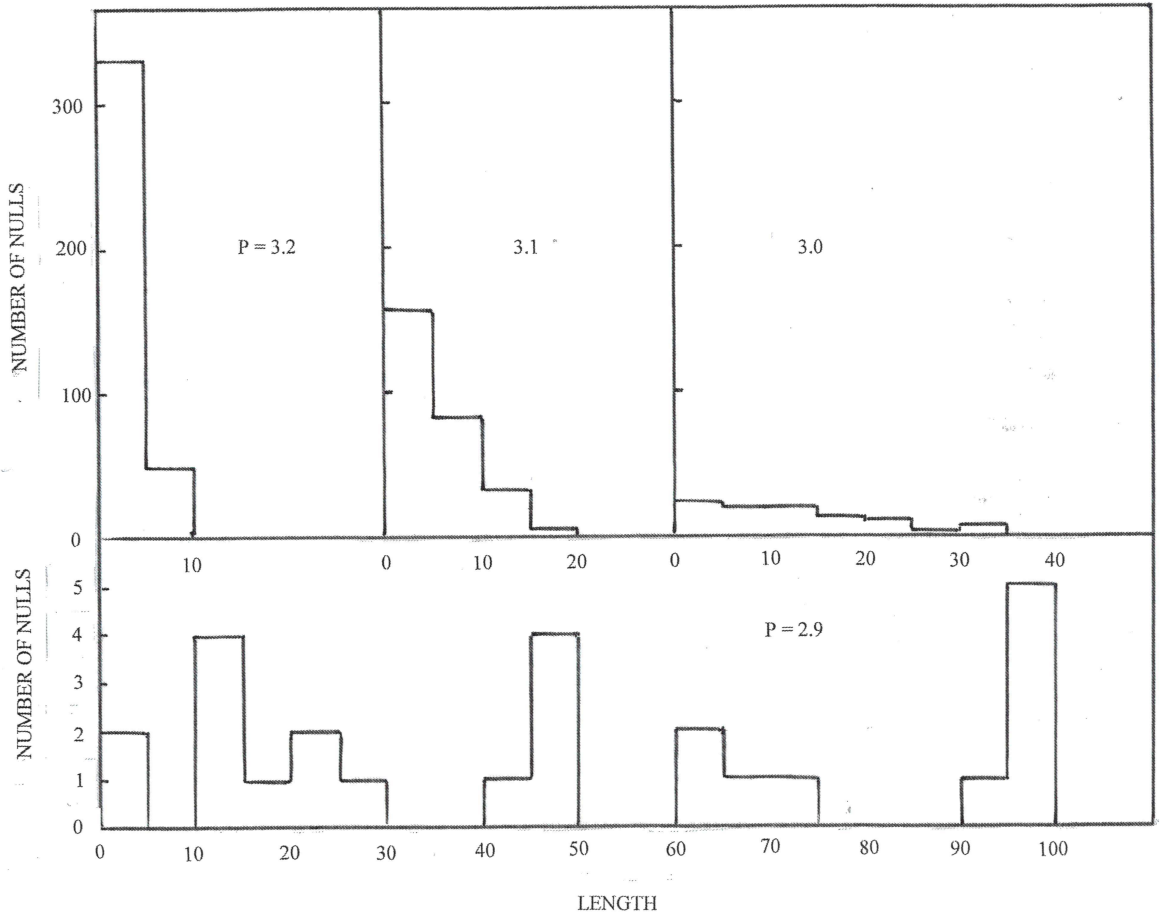}

\caption{Histograms of null length (in steps) are shown for the cases listed in Table 2.  The upper left-hand abscissa refers to $P = 3.0$, $3.1$ and $3.2$ s; the lower to $P = 2.9$ s. The ordinates have a common length scale in numbers of steps.}

\end{figure*}

\section{Exceptional pulsars}

Certain pulsars have in extreme old age values of $T_{s}$ that cannot support photo-ionization.  This class certainly includes PSR J2144-3933 first observed by Young, Manchester \& Johnston (1999) whose surface temperature is known to be $T_{s} < 4.2\times 10^{4}$ K (Guillot et al 2019) and may also include many MSP.

J2144-3933 is remarkable in having $P = 8.5$ s for which equations (3)-(5) give a very small potential, ${\rm e}\Phi_{ref} = 35$ GeV.  This is beneficial for Langmuir-mode growth rates, though the critical question is about the source of protons.  Growth rates are only slowly-varying functions of the ion-proton number density ratio, but two plasma components with different charge-to-mass ratios are essential.

More recently, a further very slowly rotating pulsar, J0250+5854, has been found by Tan et al (2018) with $P = 23.5$ s.  But its field is $B_{12} = 26$ with the significant parameter ${\rm e}\Phi_{ref} = 59$ GeV, so that it is dynamically more active than J2144-3933.

The polar-cap blackbody field in normal pulsars contributes only a small and relatively unimportant component $\epsilon_{h}$ to the reverse-electron energy flux, but for J2144-3933 it is decisive.  One-dimensional solutions of Poisson's equation
can be found at very low altitudes.  Here, the  non-relativistic stage of ion acceleration (Michel 1974) produces a component,
\begin{eqnarray}
E_{\parallel} = \left(\frac{8\pi\sigma_{GJ}(h)Am_{p}c^{2}}{\tilde{Z}{\rm e}}\right)^{1/2} = 5.0\times 10^{4}\left(\frac{AB_{12}}{P\tilde{Z}}\right)^{1/2}
\end{eqnarray}
in units of {\rm esu} within $\sim 10^{2}$ cm of the surface.  Assume $Z = 20$ as appropriate for small $\alpha_{p}$ and that photo-ionization from the atmosphere mean ion charge $\tilde{Z} \rightarrow Z_{h}$ occurs at an altitude $h_{in}$ where the reverse-electron charge reduces $E_{\parallel}$ to zero.  Then,
\begin{eqnarray}
 h_{in} = \frac{E_{\parallel}Z_{h}}{8\pi(Z_{h} - \tilde{Z})\sigma_{GJ}(h)}.
\end{eqnarray}
The maximum ion energy is at $h_{in}$ and equals ${\rm e}E_{\parallel}h_{in}/2 = 19$ GeV per unit positive charge.  Assuming photo-ionization occurs at this altitude, the value of $\epsilon_{h}$ would be,
\begin{eqnarray}
\epsilon_{h} = \frac{1}{2}{\rm e}E_{\parallel} h_{in}(Z_{h} - \tilde{Z}) =
 \frac{1}{2}Am_{p}c^{2}\frac{Z_{h}}{\tilde{Z}}.
 \end{eqnarray}
 But this is subject to the requirement that $h_{in} \ll u_{0}(1)$ and the one-dimensional approximation fails if $Z_{h} - \tilde{Z} \ll 1$ .  Thus equation (16) gives no more than an optimistic upper limit which fails at low $T_{pc}$.   Computation of photo-ionization assuming a constant value of $19$ GeV per unit positive charge at all altitudes in the polar-cap field gives $Z_{h} - \tilde{Z} = 0.4$ and $\epsilon_{h} = 12$ GeV at an assumed $T_{pc} = 5\times10^{5}$ K, which is equivalent to $2.4$ protons per ion.
But the actual polar-cap temperature given by this input alone would be
\begin{eqnarray}
 T_{pc} = \left(\frac{B\alpha_{p}}{\sigma{\rm e}PW_{p}}\right)^{1/4}
  =5.2\times 10^{5}\alpha_{p}^{1/4}  {\rm K},
\end{eqnarray}
in which $\sigma$ is here Stefan's constant. This temperature would be only very marginally adequate for a satisfactory Langmuir-mode growth rate.  But a current sheet immediately external to the surface of radius $u_{0}$ carrying a reverse particle flux is present in recent computational modelling of the outer magnetosphere (see, for example, Bai \& Spitkovsky 2010) and is likely to create a blackbody source.  The effective value of $T_{pc}$ is then increased above that of equation (17).  Our belief is that small areas of high temperature found by X-ray observations are not in the open polar-cap but are external to its boundary and associated with the reverse-flow current sheet.
For example, equation (17) gives the low temperature of $T_{pc} = 0.73\times 10^{6} \alpha_{p}^{1/4}$ K for B0950+08, which would be reduced by red-shift to the observer frame.  An enhanced blackbody field at the polar cap is likely to be a feature of all
pulsars and is likely to be important in the functioning of the proton source in some elderly pulsars.

Pulsar functioning of the two slow rotators is not surprising, but their presence completely isolated in the logarithmic $P - \dot{P}$ plane from the main group of pulsars is interesting.  It can be understood by considering the state of a neutron star with negligible $T_{s}$ but a value of ${\rm e}\Phi_{ref} \approx 10^{3}$ GeV typical of the main group.  Low altitude acceleration exists as described above but protons and ions have no interaction with blackbody radiation at altitudes $\gg u_{0}$.  There is no screening and therefore acceleration to $\gamma > \gamma_{c}$ occurs before Langmuir-mode growth is possible.  It is the two slow rotators' low values of ${\rm e}\Phi_{ref}$ that make growth possible.

Many MSP may fall into the negligible $T_{s}$ class.  Typical values $P = 3\times 10^{-3}$ s and $B = 3\times 10^{8}$ G give ${\rm e}\Phi_{ref} \approx 4\times 10^{4}$ GeV.  Equations (14) and (15) show that reverse-flux electron energies are likely to be of the order $\sim 60$ GeV.  Adequate proton production (the parameter $\tilde{K}$) occurs but screening of the field appears doubtful at altitudes $h \gg u_{0}$, where Langmuir-mode growth is expected.  It is possible that the existence of observable polar-cap emission in the MSP is a consequence of magnetospheric structural difference caused by MSP $u_{0}$ being more than an order of magnitude larger than in the main group of normal pulsars; but it is not understood.

\section{Stability and adaptability}

The stability of the states calculated in Section 3 can be disturbed by several factors.  We have earlier noted that the surface mean nuclear charge $\langle Z \rangle$ can be unstable on time-scales of the order of $\tau_{rl}$. It is likely that this form of instability is relevant to the existence of nulls longer than those illustrated in Fig. 1.  It is also significant because $W_{p}$ is a function of $Z$ (see Jones 2010).  If $Z$ values were so small in any interval of time that ions in the surface atmosphere were completely stripped of electrons so that $\tilde{Z} = Z$ there would be no reverse-electron flux, no screening of $\Phi_{ref}$ and hence in the general case, no radio emission.

A previous paper (Jones 2018) noted that the giant dipole state is also a source of neutrons at an approximately equal rate to that of protons which, through capture $\gamma$-rays, would be a source of pair creation above the polar cap in neutron stars with suitable magnetic field.  This indicated the possibility of a self-sustaining inverse Compton scattering process which would be a further mechanism for nulls and mode-changes.  But rates are very difficult to estimate with any reliability.

In Section 4, a likely contribution to the polar-cap blackbody field from a return current sheet at the open-closed boundary was described.  If it exists, it must also be a source of neutrons and a further source of pairs in the open magnetosphere above the polar cap.  This contribution  to electron-photon shower formation and the proton flux would always be present, but dependent only on the $\Phi_{i}$.  To summarize, the complexity of the system appears unlimited but not at variance with observed pulsar phenomena.

A feature of the model is that it involves a number of parameters that are not well known, specifically, the nuclear charge $Z_{0}$ immediately below the surface and the temperature $T_{s}$ of that part of the surface subtending about a steradian that is the source of the radiation through which ions are accelerated.  Ionization energies and photo-electric cross-sections are not known and have been no more than estimated at fields of the order of $10^{12}$ G.  But pulsar functioning is widely believed to be fairly common in the galactic neutron-star population which suggests that extreme values of parameters, such as the flux-line curvatures needed to produce high-density electron-positron plasmas, are not a suitable basis for any model.

That featured here is parameter-adaptable.  In cooling, the surface must pass through the band of $T_{s}$ that is required for the observed phenomena.  It is also the case , particularly at high $B$, that electron photo-ionization thresholds decrease adiabatically as the altitude increases. Cross-sections though only estimated, are large near thresholds and it is certain that ions initially in local thermodynamic equilibrium in the neutron-star atmosphere must at some acceleration Lorentz factor undergo photo-ionization.  The acceleration potential induced by the Lense-Thirring effect is generous and except in cases such as J2144-3933 adequate to produce the necessary Lorentz factors.  With regard to the MSP, the physics of proton formation at $B_{12} \ll 1$ is unchanged except that Rydberg ionization energies are used and zero-field photo-ionization cross-sections are needed: also shower structure is not qualitatively changed.

Two phenomena, not considered here, are mode-changes and sub-pulse drift. The latter is particularly complex: in certain cases both directions of drift, one in each conal component, are observed within a single pulse.  This is difficult to reconcile with ${\bf E}\times{\bf B}$ motion above the polar cap.  Mode-changes have been considered in a previous publication (Jones 2018).  It may be possible to show that sub-pulse drift can be maintained in a linear and quasi-stable manner starting from a suitable initial state, but this has not yet been achieved.

\section*{Acknowledgments}

The author wishes to thank the anonymous referee for some important guidance on the presentation of this work.

\bsp

\label{lastpage}

\end{document}